\documentclass[aps,pra,twocolumn,superscriptaddress,amsmath,amssymb,tightenlines,epsfig,floatfix]{revtex4}
\pdfoutput=1
\usepackage{graphicx}
\usepackage{dcolumn}	
\usepackage{bm}			
\usepackage{amsfonts}
\usepackage{xspace}

\begin{document}

\newcommand{\Ham}{\ensuremath{\hat{\mathcal{H}}}\xspace}
\newcommand{\bra}[1]{\ensuremath{\langle #1|}\xspace}
\newcommand{\ket}[1]{\ensuremath{|#1\rangle}\xspace}
\newcommand{\ddt}[1]{\ensuremath{\frac{d #1}{d t}}\xspace}
\newcommand{\psihat}{\ensuremath{\hat{\psi}}\xspace}
\newcommand{\psihatd}{\ensuremath{\hat{\psi}^{\dagger}}\xspace}
\newcommand{\Ehat}{\ensuremath{\hat{E}}\xspace}
\newcommand{\Ehatd}{\ensuremath{\hat{E}^{\dagger}}\xspace}
\newcommand{\ahat}{\ensuremath{\hat{a}}\xspace}
\newcommand{\ahatd}{\ensuremath{\hat{a}^{\dagger}}\xspace}
\newcommand{\bhat}{\ensuremath{\hat{b}}\xspace}
\newcommand{\chat}{\ensuremath{\hat{c}}\xspace}
\newcommand{\Vhat}{\ensuremath{\hat{V}}\xspace}
\newcommand{\bhatd}{\ensuremath{\hat{b}^{\dagger}}\xspace}
\newcommand{\chatd}{\ensuremath{\hat{c}^{\dagger}}\xspace}
\newcommand{\boldr}{\ensuremath{\mathbf{r}}\xspace}
\newcommand{\dr}{\ensuremath{\,d^3\mathbf{r}}\xspace}
\newcommand{\expect}[1]{\ensuremath{\left< #1\right>}\xspace}
\newcommand{\etal}{\emph{et al.\/}\xspace}
\newcommand{\ie}{i.e.\:}
\newcommand{\rf}{R.F.\,\xspace}
\newcommand{\half}{\ensuremath{\frac{1}{2}}\xspace}
\newcommand{\phivector}{\ensuremath{\begin{bmatrix}\phi_+(z) \\ \phi_{-}(z)\end{bmatrix}}\xspace}
\newcommand{\psivector}{\ensuremath{\begin{bmatrix}\psi_t(z) \\ \psi_u(z)\end{bmatrix}}\xspace}
\newcommand{\talph}{\ensuremath{\tilde{\alpha}}\xspace}
\newcommand{\eq}[1]{Eq.\,(\ref{#1})\xspace}
\newcommand{\fig}[1]{Figure\,(\ref{#1})\xspace}
\newcommand{\vs}[1]{\ensuremath{\boldsymbol{#1}}\xspace}
\renewcommand{\v}[1]{\ensuremath{\mathbf{#1}}\xspace}
\newcommand{\Psihat}{\ensuremath{\hat{\Psi}}\xspace}
\newcommand{\Psihatd}{\ensuremath{\hat{\Psi}^{\dagger}}\xspace}
\newcommand{\Vhatd}{\ensuremath{\hat{V}^{\dagger}}\xspace}
\newcommand{\Xhat}{\ensuremath{\hat{X}}\xspace}
\newcommand{\Xhatd}{\ensuremath{\hat{X}^{\dag}}\xspace}
\newcommand{\Yhat}{\ensuremath{\hat{Y}}\xspace}
\newcommand{\Yhatd}{\ensuremath{\hat{Y}^{\dag}}\xspace}
\newcommand{\Nhat}{\ensuremath{\hat{N}}\xspace}
\newcommand{\oppy}{\ensuremath{\hat{\mathcal{O}}}\xspace}

\title{A dynamic scheme for generating number squeezing in Bose-Einstein condensates through nonlinear interactions}
\author{Simon A. Haine}
\affiliation{Australian Research Council Centre of Excellence for Quantum-Atom Optics}
\affiliation{School of Physical Sciences, University of Queensland, Brisbane, 4072, Australia}
\email{haine@physics.uq.edu.au}

\author{Mattias T. Johnsson}
\affiliation{Australian Research Council Centre of Excellence for Quantum-Atom Optics}
\affiliation{Department of Quantum Science, The Australian National University, Canberra, 0200, Australia}
\preprint{Version: submission \today}

\begin{abstract}
We develop a scheme to generate number squeezing in a Bose-Einstein condensate by utilizing interference between two hyperfine levels and nonlinear atomic interactions. We describe the scheme using a multimode quantum field model and find agreement with a simple analytic model in certain regimes. We demonstrate that the scheme gives strong squeezing for realistic choices of parameters and atomic species. The number squeezing can result in noise well below the quantum limit, even if the initial noise on the system is classical and much greater than that of a Poissonian (shot noise limit) distribution. 
\end{abstract}

\maketitle

\section{Introduction} 
The experimental realization of Bose-Einstein condensates (BECs) has allowed the creation of macroscopic quantum systems that are highly controllable, and hence provide an excellent system to test predictions of many body quantum dynamics. The generation of nonclassical states in BECs, such as number squeezed states, would allow for measurements of particle number statistics that differ from classical predictions \cite{scully_book}. However, while the motional state of the atoms is reasonably simple to manipulate, the quantum statistics governing the number distribution of the particles is difficult to control, with BECs typically produced with a large ($5\%$) shot-to-shot variation in the number. The generation of nonclassical states in BECs is currently of great interest \cite{oberthaler_nature}, as it could potentially enhance the sensitivity of atomic interferometers \cite{dowling} used to measure electric, magnetic, and gravitational fields, accelerations, and atomic interactions. The generation of nonclassical states also provides a method for testing the fidelity of recent quantum state transfer schemes \cite{bohh_tele}, and an atom laser produced from a number squeezed BEC will have a reduced linewidth \cite{mattias_qlinewidth}. 

The generation of nonclassical states via self-interaction in samples of cold atoms has been considered before \cite{wuster, kerrsqueezy1, sorensen, dutton, sinatra, nandi}. The schemes proposed in \cite{wuster} and \cite{kerrsqueezy1} describe the generation of quadrature squeezing, which requires the use of a well-defined phase reference in order to be observed. A well-defined phase reference is difficult to obtain in atom optical systems, especially in the presence of strong nonlinearities, which are required to produce the quadrature squeezing, thus making the schemes somewhat unrealistic. In addition, both of these schemes assume the BEC is initially in a coherent state, rather than a more realistic statistical mixture of coherent states with random phases. \cite{sorensen, dutton, sinatra} have demonstrated that the atomic nonlinearity can be used to generate number difference squeezing, creating a state with angular momentum projection below the standard quantum limit (spin squeezing). These schemes assume that the total number of particles is initially well defined. Recently, Esteve \etal \cite{oberthaler_nature} have directly observed squeezing in the number difference between two adjacent lattice sites. 

Chuu \etal \cite{raizen_prl} have demonstrated the ability to produce small condensates which exhibit number squeezing. By producing a very stable trapping potential, they found that they were able to produce condensates with a very well specified chemical potential. As the chemical potential is related to the condensate number through the nonlinear interaction, this leads to number squeezed condensates. 

In this article we describe a scheme that allows the creation of absolute number squeezing in a BEC (as opposed to number difference squeezing), which is experimentally realistic, utilizes only the relatively simple experimental technique of Ramsey interferometery, and does not require manipulation of the scattering length via a Feshbach resonance or a coherent phase reference for the atoms. We show that our scheme achieves number squeezing below the quantum limit even if there is initially considerable classical noise on the number statistics, and that this result holds even when we realistically assume that the initial state of the BEC is a statistical mixture of states with random phase.

We consider a BEC with two internal states confined to an optical trap, with all the atoms initially in one state. A short state-dependent coupling is applied, transferring a small fraction of the population to another state. The system is then left to evolve for some time, allowing nonlinear interactions and interference between the two states, before the coupling is applied for a second time, transferring some of the population back to the initial state. Provided the $s$-wave scattering lengths of the atoms in the different internal states are not all identical, and by choosing appropriate coupling strengths, hold times and trap geometry, it is possible to generate number squeezing. An important difference between our scheme and the schemes demonstrated in refs. \cite{oberthaler_nature} and \cite{raizen_prl} is that our scheme is based on dynamic interference between the two modes to obtain absolute number squeezing in one of the modes, where as the schemes demonstrated in refs. \cite{oberthaler_nature} and \cite{raizen_prl} obtain their squeezing by adiabatically changing the potential to one where the ground state of the system exhibits number difference squeezing (in the case of \cite{oberthaler_nature}) or absolute number squeezing (in the case of \cite{raizen_prl}).

\begin{figure}
\includegraphics[width=0.8\columnwidth]{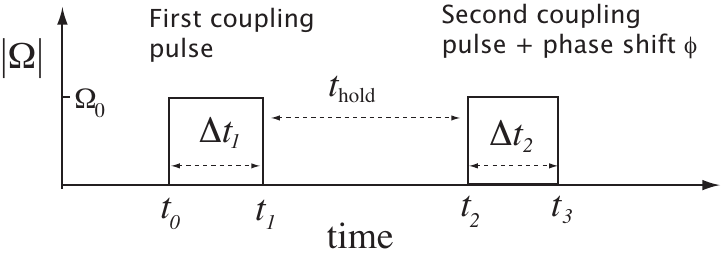}
\caption{\label{timingfig} Timing for the coupling pulses in the proposed experiment. The coupling field is turned on at $t_0$, and then off again at $t_1$. After a duration $t_{\mathrm{hold}}$, the coupling pulse is turned back on again at $t_2$ and finally turned off at $t_3$. After this, the population of state $|2\rangle$ atoms is measured.  }
\end{figure}
\section{Scheme.} Our proposed scheme is based on a Ramsey interference experiment between two hyperfine states of sodium, namely $|F=2, m_F = 0\rangle \equiv |1\rangle$ and $|F=1, m_F = +1\rangle \equiv |2\rangle$. The timing for our scheme is outlined in Figure \ref{timingfig}. We begin with all the atoms in a BEC in the $|1\rangle$ state in an optical trap. At $t = t_0$, the microwave coupling is turned on for a brief duration of time $\Delta t_1 = t_1 - t_0$. During this time, a fraction of the atoms are then transferred to state $|2\rangle$. The coupling is switched off, and the system is left to evolve for an amount of time $t_{\mathrm{hold}} = t_2-t_1$, before we interfere the two modes with a second microwave coupling pulse for a duration $\Delta t_2 = t_3-t_2$. We assume that this pulse is phase locked to the first, with an adjustable phase shift $\phi$. This phase shift is an important ingredient, as it is what allows us to ensure that the two modes interfere in such a way as to produce number squeezing, independently of $t_{\mathrm{hold}}$, which controls the {\it depth} of squeezing. Nandi \etal \cite{nandi} have recently published a scheme based on a Ramsey-Bord\'e interferometer, but without including this adjustable phase shift $\phi$ at the second coupling pulse. As a result, their squeezing is only observed at particular values of $t_{\mathrm{hold}}$, when the nonlinear phase shift acquired during the hold time is appropriate. The inclusion of this phase shift gives us an additional degree of freedom in our system, as we can choose $t_{\mathrm{hold}}$ independently of the phase shift required to produce number squeezing, and thus optimize our squeezing depth. After $t_3$, we separate the two modes with a magnetic field, and count the number of atoms in mode $|2\rangle$ to determine the number statistics. The Hamiltonian for the system is $\Ham = \Ham_0 + \Ham_c(t)$, with
\begin{eqnarray}
 \Ham_0 &=& \sum_{j=1,2}\int \psihatd_i(\boldr) H_j
 \psihat_i(\boldr) \dr \nonumber \\ 
&+& \sum_{i,j=1,2}\frac{U_{ij}}{2}\int \psihatd_i(\boldr)\psihatd_j(\boldr)\psihat_i(\boldr)\psihat_j(\boldr)\, \dr \, , \label{ham1}
\end{eqnarray}
and
\begin{eqnarray}
\Ham_c &=&  \int \left(\hbar\Omega(t)e^{i\phi} \psihatd_{2}(\boldr)\psihat_1(\boldr) + \mathrm{h.c.}\right)\, \dr \, , \label{ham2}
\end{eqnarray}
where $\psihat_i(\boldr)$ represents the annihilation operator for state $|i\rangle$, $H_j = \frac{-\hbar^2}{2m}\nabla^2 + V_{\mathrm{opt}}(\boldr) + (j-1)\hbar\delta$ is the single particle Hamiltonian, and $V_{\mathrm{opt}}(\boldr)$ is the optical dipole potential. $\Omega(t)$ represents the microwave coupling field, which is switched on and off to control the coupling between the two hyperfine levels. The phase of this RF field can also be tuned between each pulse. We will make the rotating wave approximation \cite{scully_book} and assume that the coupling is on resonance, such that $\Omega(t) = \Omega_0 e^{-i \delta t}$, with $\hbar\delta$ the hyperfine splitting between $|1\rangle$ and $|2\rangle$, and $\Omega_0$ is the Rabi frequency. 

\section{Analytic model.} 
We first consider a two mode model, which demonstrates how atomic nonlinearities can be used to generate number squeezing. A two-mode model can be derived from \eq{ham1} and \eq{ham2} by assuming that the atoms remain in the ground motional state of the optical trap. With this assumption, the modified Hamiltonian for the system is $\tilde{\mathcal{H}} = \tilde{\mathcal{H}}_0 + \tilde{\mathcal{H}}_c$, with
\begin{eqnarray}
\tilde{\mathcal{H}}_{0} &=& \hbar\delta\ahatd_2\ahat_2 + \sum_{i,j = 1,2}\hbar\chi_{ij}\ahatd_i\ahat_i\ahatd_j\ahat_j \\\tilde{\mathcal{H}}_{c} &=& \hbar\left(\Omega(t)e^{i\phi}\ahatd_2\ahat_1 + \mathrm{h.c.}\right)
\end{eqnarray}
where $\ahat_1$ ($\ahat_2$) annihilates an atom from state $|1\rangle$ ($|2\rangle$), and $\chi_{ij} = \frac{U_{ij}}{2}\int |\psi_0(\boldr)|^4 \, \dr$, where $\psi_0(\boldr)$ is the ground state wavefunction of the optical potential. 

Beginning at $t_0$, with the coupling initially switched off ($\Omega_0 = 0$), and assuming that our initial state is a Poissonian mixture of number states for mode $|1\rangle$, and vacuum for mode $|2\rangle$, the density matrix for the system is
\begin{equation}
\rho(t_0) = e^{-|\alpha_0|^2}\sum_{n_1=0}^{\infty}\frac{(|\alpha_0|^{2})^{n_1}}{n_1!}|n_1,0\rangle\langle n_1,0|
\end{equation}
where the state $|n_1, n_2\rangle$ denotes $n_1$ atoms in mode $|1\rangle$ and $n_2$ atoms in mode $|2\rangle$ and $|\alpha_0|^2 \equiv N_0$ is the mean number of atoms. We note that $\rho(t_0)$ is mathematically equivalent to a mixture of coherent states with random phases:
\begin{equation}
\rho(t_0) = \frac{1}{2\pi} \int_0^{2\pi} |\alpha_0 e^{i\theta}\rangle\langle\alpha_0 e^{i\theta}| \, d\theta \otimes|0\rangle\langle0|
\end{equation}
where $|\alpha\rangle \equiv e^{-|\alpha|^2/2}\sum_{n}\frac{\alpha^n}{\sqrt{n!}}|n\rangle$ denotes the Glauber coherent state \cite{scully_book}. At times $t<t_0$, the evolution is trivial as $\rho$ commutes with $\tilde{\mathcal{H}}_0$.

At $t_0$, the coupling is turned on for a duration $\Delta t_1$ coupling a fraction of the atoms ($\sin^2(\Omega_0 \Delta t_1)$) into mode $|2\rangle$. If $\Delta t_1$ is sufficiently short that we can ignore the contribution to the evolution due to the nonlinear part of  $\tilde{\mathcal{H}}_0$,  the density matrix for the system becomes
\begin{eqnarray}
\rho(t_1) &=& \int_0^{2\pi} |\alpha(t_1) e^{i\theta}\rangle\langle \alpha(t_1) e^{i\theta}| \otimes |\beta(t_1) e^{i\theta}\rangle\langle \beta(t_1) e^{i\theta}| \, \frac{d\theta}{2\pi} \nonumber \\
&=& \frac{e^{-|\alpha|^2-|\beta|^2}}{2\pi}\sum_{n_1,n_2, m_1, m_2} \int_0^{2\pi}e^{i\phi (m_1+m_2 -n_1 -n_2)} d\phi \nonumber \\
&&\frac{\alpha(t)^{n_1}\alpha^*(t)^{m_1}\beta(t)^{n_2}\beta^*(t)^{m_2}}{\sqrt{n_1!m_1!n_2!m_2!}} |n_1, n_2\rangle\langle m_1, m_2 | \nonumber \\ 
&=& \sum_{n_1,m_1, n_2} A_{n_1, m_1, n_2} |n_1, n_2\rangle\langle m_1, n_1+n_2-m_1 | \, ,
\end{eqnarray}
with
\begin{eqnarray}
A_{n_1, m_1, n_2} &=& e^{-|\alpha(t_1)|^2-|\beta(t_1)|^2}  \\ 
&\times&\frac{\alpha(t_1)^{n_1}\alpha^*(t_1)^{m_1}\beta(t_1)^{n_2}\beta^*(t_1)^{n_1+n_2-m_1}}{\sqrt{n_1!m_1!n_2!(n_1+n_2-m_1)!}}  \nonumber \, , 
\end{eqnarray} 
and  $\alpha(t_1) = \alpha_0\cos \theta_1$, $\beta(t_1) = - i\alpha_0\sin \theta_1$, where $\theta_1 = \Omega_0\Delta t_1$. We note that although the condensate initially had no global phase, a relative phase between the two modes has been created by this first coupling pulse.

The coupling is now switched off, and the system is left to evolve under $\tilde{\mathcal{H}}_0$ for a period of time $t_{\mathrm{hold}}$. At $t=t_2$, the density matrix for the system is now
\begin{eqnarray}
\rho(t_2) &=& \sum_{n_1, n_2, m_1 = 0}^{\infty} A_{n_1, n_2, m_1}  e^{-i \Phi_{n_1, n_2, m_1}t_{\mathrm{hold}}} \nonumber \\ 
&\times& |n_1, n_2\rangle\langle m_1, n_1\,+n_2\,-m_1| \, ,
\end{eqnarray}
with 
\begin{eqnarray}
\Phi_{n_1, n_2, m_1} &=& \chi_{11}(n_1(n_1-1) -m_1(m_1-1)) \nonumber \\ 
&+& \chi_{22}((m1-n1)(2n_2 + n_1 - m_1 -1)) \nonumber \\
 &+&  2\chi_{12}(n_1 n_2 - m_1(n_1+n_2-m_1)) \nonumber \\
 &+& \delta(m_1 -n_1) \, .
\end{eqnarray}
At this point, both modes still contain a Poissonian number distribution, but the relative phase created in the previous step has been `sheared' due to the nonlinear interaction. We note that if $\chi_{11}=\chi_{22} = \chi_{12}$, there is no phase shearing due to this effect, and our scheme does not work. We chose sodium as our atomic species, as it has a relatively large difference between the scattering lengths of $|F=1, m_F = +1\rangle$ and $|F=2, m_F=0 \rangle$.

\begin{figure}
\includegraphics[width=1.0\columnwidth, bb= 0 0 800 800,clip]{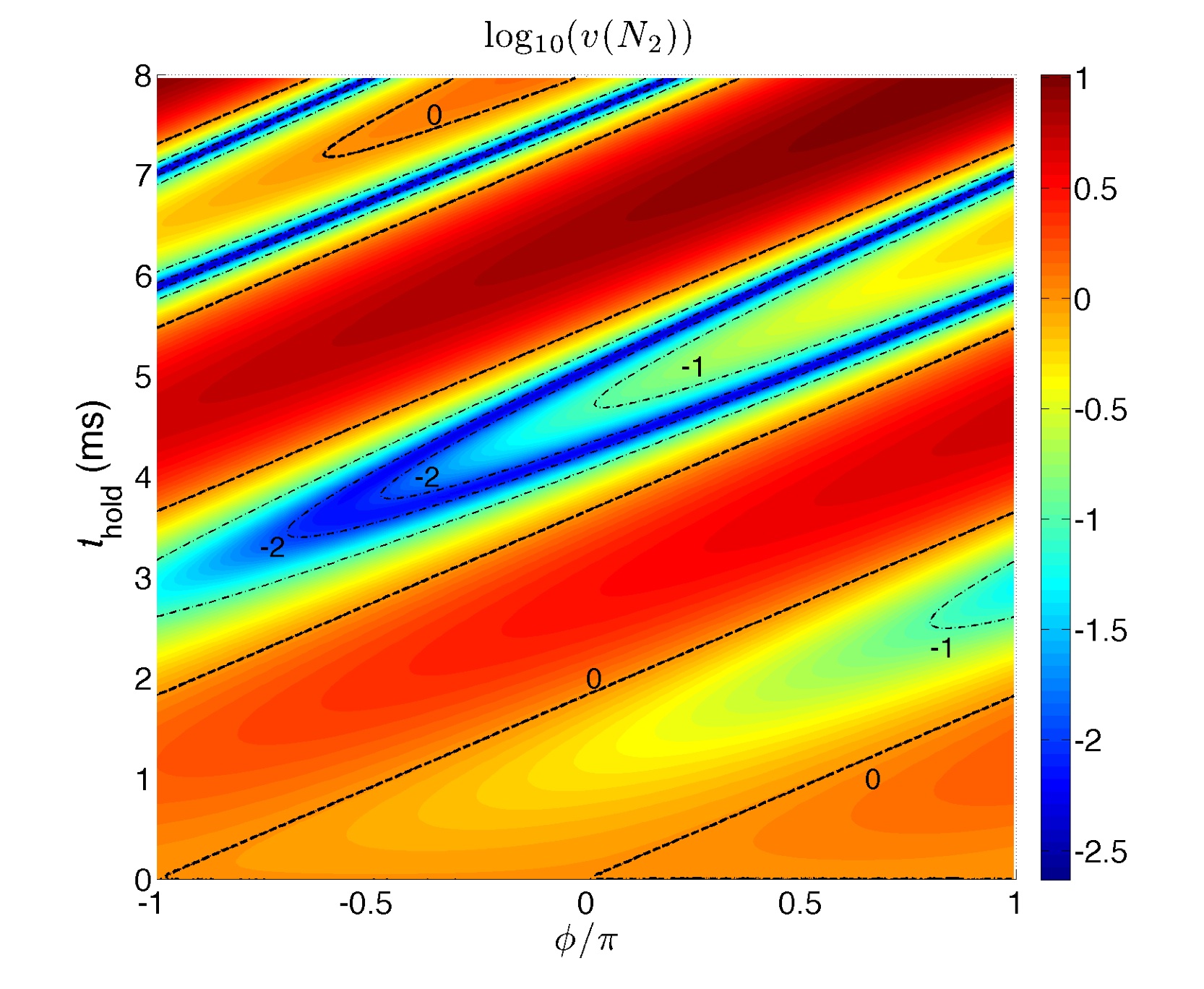}
\caption{\label{holdscan} (color online) $\log_{10}(v(\Nhat_2))$ as a function of $t_{\mathrm{hold}}$ and $\phi$. For some values of $t_{\mathrm{hold}} $ and $\phi$, $v(\Nhat_2)$ dips below $0.01$, indicating significant squeezing. Parameters: $\chi_{11} = \chi_{12} = 0.018$ s$^{-1}$, $\chi_{22} = 0.019$ s$^{-1}$. The strength of the coupling pulses was $\theta_1 =0.3$ rad and $\theta_2 = 0.025$ rad for the first and second coupling pulses respectively. The initial occupation of the BEC was chosen to be a Poisson distribution with $\langle \Nhat\rangle = 10^7$. These parameters correspond to the scattering properties of sodium in a 500 Hz spherical harmonic trap.   }
\end{figure}
  Finally, we describe the dynamics caused by the second microwave pulse in the Heisenberg picture, by noting that the Heisenberg operators after the second pulse are $\ahat_{1H} = \ahat_1(0)\cos{\theta_2} -i\ahat_2(0)\sin{\theta_2} e^{i\phi}$, and $\ahat_{2H} = \ahat_2(0)\cos{\theta_2} - i\ahat_1(0)\sin{\theta_2} e^{-i\phi}$, with $\theta_2 = \Omega_0 \Delta t_2$, where $\Delta t_2$ is the duration of the second microwave pulse, and $\phi$ is the phase of the microwave field relative to the first pulse. Again, we have assumed that the duration of the pulse is sufficiently short that we can ignore the evolution due to the nonlinear component of $\tilde{\mathcal{H}}_0$. As we are only interested in the number statistics, we can neglect the rest of the evolution after the second microwave pulse, as the number operators for both modes commute with $\tilde{\mathcal{H}}_0$. Assuming we can distinguish state $|2\rangle$ atoms from state $|1\rangle$ atoms, we define the normalized number variance for state $|2\rangle$ atoms as $v(\Nhat_2) \equiv  (\langle \Nhat_2^2\rangle - \langle \Nhat_2\rangle^2)/\langle \Nhat_2 \rangle$, $\Nhat_2 \equiv \ahatd_{2H}\ahat_{2H}$. Figure \ref{holdscan} shows the $v(\Nhat_2)$ as a function of $t_{\mathrm{hold}}$ and $\phi$, using the scattering properties of sodium ($a_{11} = a_{12} = 2.8$ nm, $a_{22} = 3.0$ nm \cite{samuelisET2000} in a $500$ Hz spherical harmonic trap. For some values of $t_{\mathrm{hold}} $ and $\phi$, $v(\Nhat_2)$ dips below $0.01$, as compared to the quantum limit $v(\Nhat_2) = 1$ associated with a coherent state, indicating significant number squeezing. The parameter space for this model is quite large, as we can adjust the length of the first and second coupling pulses, the hold time, and the phase of the second coupling pulse. If we did not have the ability to adjust the phase of the second coupling pulse, we would be constrained to a vertical line in Figure \ref{holdscan}, and not necessarily be able to access the optimum value of the squeezing.  We found that the best number squeezing was obtained when the first coupling pulse was quite weak (approximately $8 \%$ of the atoms transferred). We also found we could still get a good level of squeezing when we began with an initial state which had number fluctuations $150$ times larger than a Poissonian distribution (about $5\%$ shot to shot fluctuations in the number) (see Figure \ref{variancefig}). When starting with a such an initial condition, the best squeezing was found when the first beam splitter was relatively weak. This is due to the fact that the addition of vacuum to a super-Poissonian number distribution drives it towards a Poissonian number distribution.

\section{Multimode model.}
To investigate if the approximations we made in the previous section to obtain an analytic solution were valid, we performed a 1D multimode simulation of the system using a stochastic phase space method. Specifically, we utilize a Truncated Wigner (TW) approach \cite{wigner_method}. We reduce Eqs. (\ref{ham1}) and (\ref{ham2}) to one dimension by integrating out the dynamics in the $y$ and $z$ dimensions. A Fokker Plank equation (FPE) is then found from the master equation for the system using the Wigner representation. This equation can then be converted into a set of stochastic partial differential equations (SPDEs), which can be solved numerically. By averaging over many trajectories with different noises, expectation values of quantities corresponding to operators in the full quantum field theory can be extracted. When converting our FPE to a SPDEs, we ignore third and higher order derivatives in the FPE, as these terms  do not have a simple mapping to the stochastic PDEs, and can be assumed to be negligible when the field has a high occupation number \cite{wigner_method}. This truncated Wigner approximation will eventually fail, as it can not describe negative components of the Wigner function, which eventually occur when evolving under a Hamiltonian such as \eq{ham1}. However, we have checked our simulations in limits where the multimode dynamics can be neglected and they agree with the results obtained from our two-mode analytical model. In addition, over our simulation times no anomalous results such as signficant negative densities were seen, indicating that the truncation of third order derivatives was a valid approximation. The SPDEs describing the one dimensional system are
\begin{eqnarray}
i\hbar\dot{\psi}_1(x) &=& \mathcal{L}_1\psi_1(x) + \hbar\Omega(t)\psi_2(x)  \\
i\hbar\dot{\psi}_2(x) &=& \mathcal{L}_2\psi_2(x) + \hbar\Omega^*(t)\psi_1(x),  
\end{eqnarray}
with
\begin{equation}
\mathcal{L}_j = H_j + U_{jj}(|\psi_j|^2-1/dx) + U_{ij}(|\psi_i|^2-1/(2dx)) \, ,
\end{equation}
where $dx$ is the grid spacing of the numerical simulations. The terms inversely proportional to $dx$ compensate for the mean field of the vacuum, which is nonzero in the Wigner approach. The noise on the initial conditions for each trajectory of the evolution of these equations was chosen such that they corresponded to the specific initial state of interest.

\begin{figure}
\includegraphics[width=0.7\columnwidth]{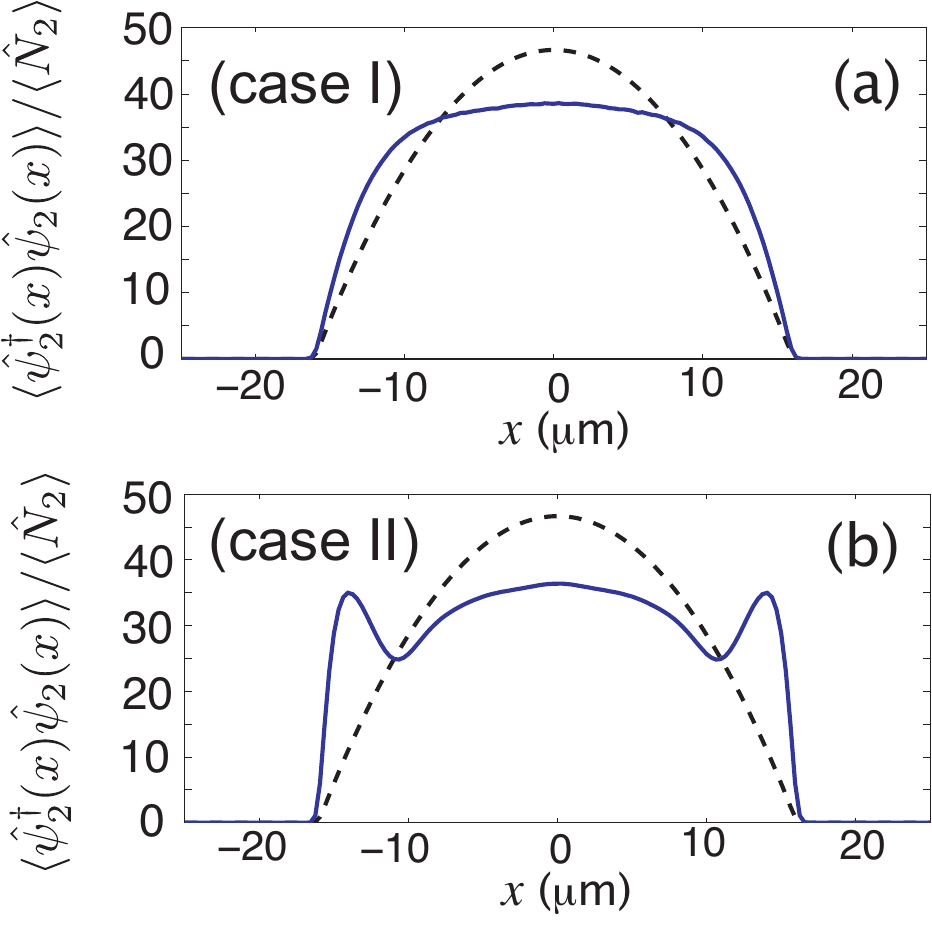}
\caption{\label{densfigs} (color online) Normalized density profile $\langle \psihatd_2(x) \psihat_2(x)\rangle / \langle \Nhat_2\rangle$ at $t_2$ (blue solid line), compared to the normalized density profile at $t_1$ (black dashed line), as calculated by the TW model for two different parameter regimes. In case II, more atoms are transferred in the first coupling pulse, which creates significant multimode dynamics. In case I, the dynamics are much less pronounced and the density profile at $t_2$ differs only slightly from that at $t_1$. Parameters: Case I: $\Delta t_1 = 1$ $\mu$s, $\Delta t_2 = 0.5$ $\mu$s, $t_{\mathrm{hold}} =  16$ ms. Case II:  $\Delta t_1 = 6$ $\mu$s, $\Delta t_2 = 0.5$ $\mu$s, $t_{\mathrm{hold}} =  3.8$ ms. In both cases we assumed $\Omega_0 = 50$ rad s$^{-1}$, and a $500$ Hz spherical harmonic trap. We assumed an initial number distribution which was poissonian, with a mean number of atoms $N_0 = 10^7$.}
\end{figure}

\begin{figure}
\includegraphics[width=1.0\columnwidth]{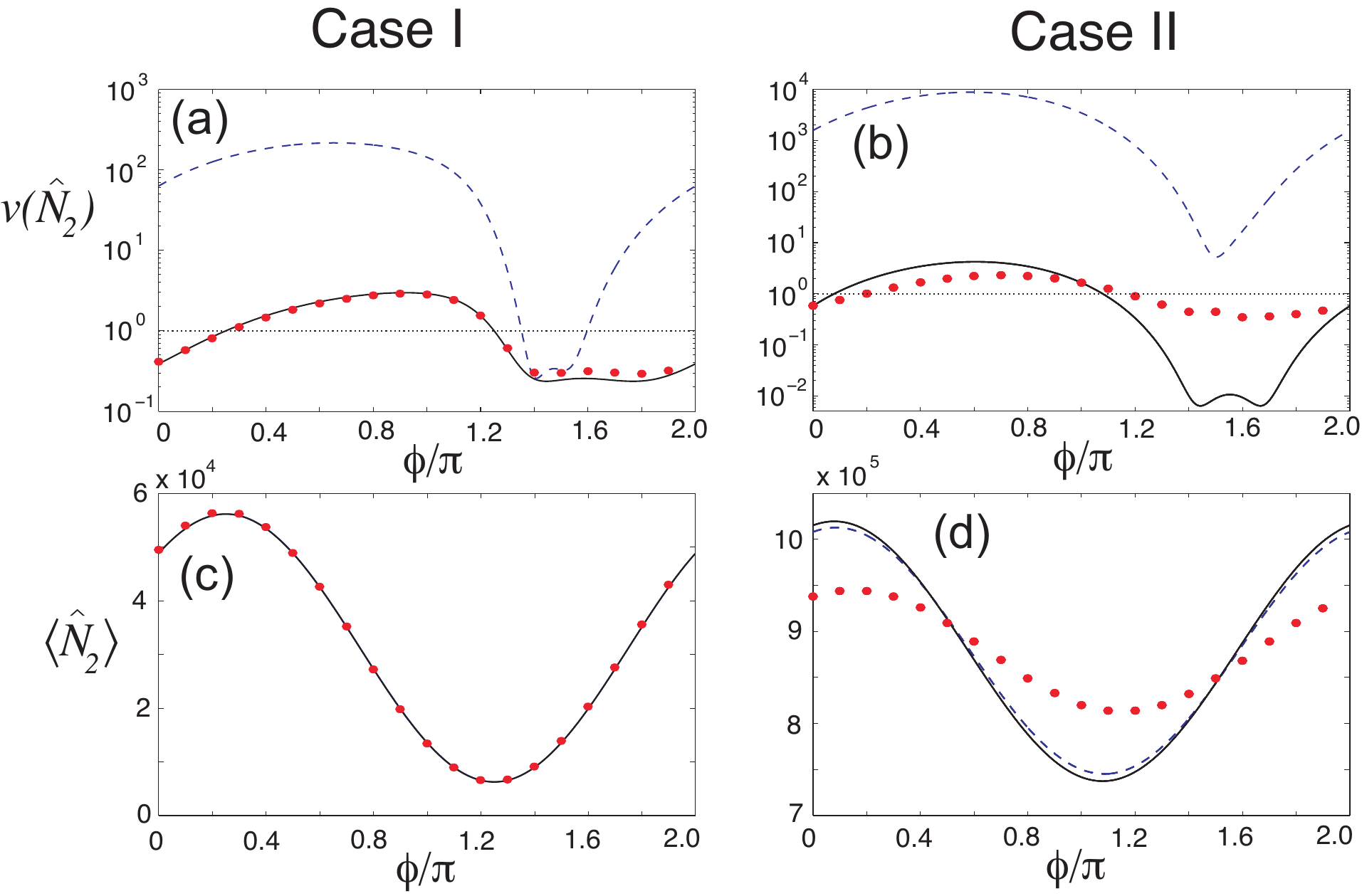}
\caption{\label{variancefig} (color online)
Results from the multimode TW model (red dots) compared to the analytic two mode model (black trace), for two different parameter regimes (case I (left column) and case II (right column) respectively). (a) and (b) show number variance $v(\Nhat_2)$, while (c) and (d) show $\langle \Nhat_2\rangle$, at $t_3$. In both cases we assumed an initial number distribution which was Poissonian, with a mean number of atoms $N_0 = 10^7$. In case I ((a) and (c)), there is excellent agreement between the multimode TW model and the two mode model. In case II ((b) and (d)), there is significant disagreement between the two models because a larger fraction of atoms is transferred during the first coupling pulse, which creates significant multimode dynamics, as can be seen in Figure (\ref{densfigs}(a)).  The blue dashed trace indicated results from the two mode model, when a superpoissonian  distribution was used at the initial state, with $5\%$ number uncertainty (approximately 150 times noisier than a Poisson distribution). (a) shows that even with large amounts of classical noise, it is possible to number squeeze below the quantum limit. Parameters: Case I: $\Delta t_1 = 1$ $\mu$s, $\Delta t_2 = 0.5$ $\mu$s, $t_{\mathrm{hold}} =  16$ ms. Case II:  $\Delta t_1 = 6$ $\mu$s, $\Delta t_2 = 0.5$ $\mu$s, $t_{\mathrm{hold}} =  3.8$ ms. In both cases we assumed $\Omega_0 = 50$ rad s$^{-1}$, and a $500$ Hz spherical harmonic trap.}
\end{figure}

Figure (\ref{densfigs}) shows the normalized density profile of state $|2\rangle$ atoms, $\langle \psihatd_2(x)\psihat_2(x)\rangle / \langle \Nhat_2\rangle$, at $t_2$, compared to the normalized density profile at $t_1$, as calculated by the TW model for two different parameter regimes (case I and case II). Case II shows an example where the multimode dynamics is significant, and the density profile at $t_2$ has a pronounced difference from the ground state density profile. The multimode dynamics is a consequence of the unequal scattering lengths, meaning that when atoms are created in state $|2\rangle$, they are no longer in a motional eigenstate of the system. These dynamics are relatively insignificant in case I, when only $0.25 \%$ of the atoms are transferred in the first coupling pulse. However, in the case II, $\sim 9\%$ of the atoms are transferred in the first coupling pulse, and the perturbation to the dynamics during the hold time is significant, even though the system is left to evolve for a much shorter time. Figure (\ref{variancefig}) shows the number of atoms in state $|2\rangle$ and the variance in the number, after the second coupling pulse, for case I and case II. In case I there is excellent agreement between the multimode model TW and the two mode analytic model. However, in case II the comparison between the two models is poor, due to significant multimode dynamics (which can be seen in Figure (\ref{densfigs}(b)) preventing the system acting as a two mode system.

\section{Experimental Considerations}
This scheme relies on having good control of microwave fields in order to implement the precise timing and resonance conditions. Precise control of microwave intensity and pulse duration is routinely achievable in atom optics laboratories (see, for example \cite{cornell_microwave}), and control of microwave frequencies with sub-Hertz stability (much less than the Fourier width of the pulses in our scheme) is routinely achievable with off the shelf equipment. As the parameter space for the experiment is large, a good knowledge of the parameters such as the trapping frequency, the Rabi frequency, and the scattering lengths will be required such that theoretical modeling can predict roughly were to seach for the squeezing. The values of the scattering lengths will probably be the least well known of these quantities. To simulate the effect of an imprecise knowledge of the scattering lengths, we have investigated the effect of varying one of the scattering lengths ($a_{22}$), while keeping all other parameters fixed to the values used in Figure (\ref{variancefig}) case I. 

\begin{figure}
\includegraphics[width=0.8\columnwidth]{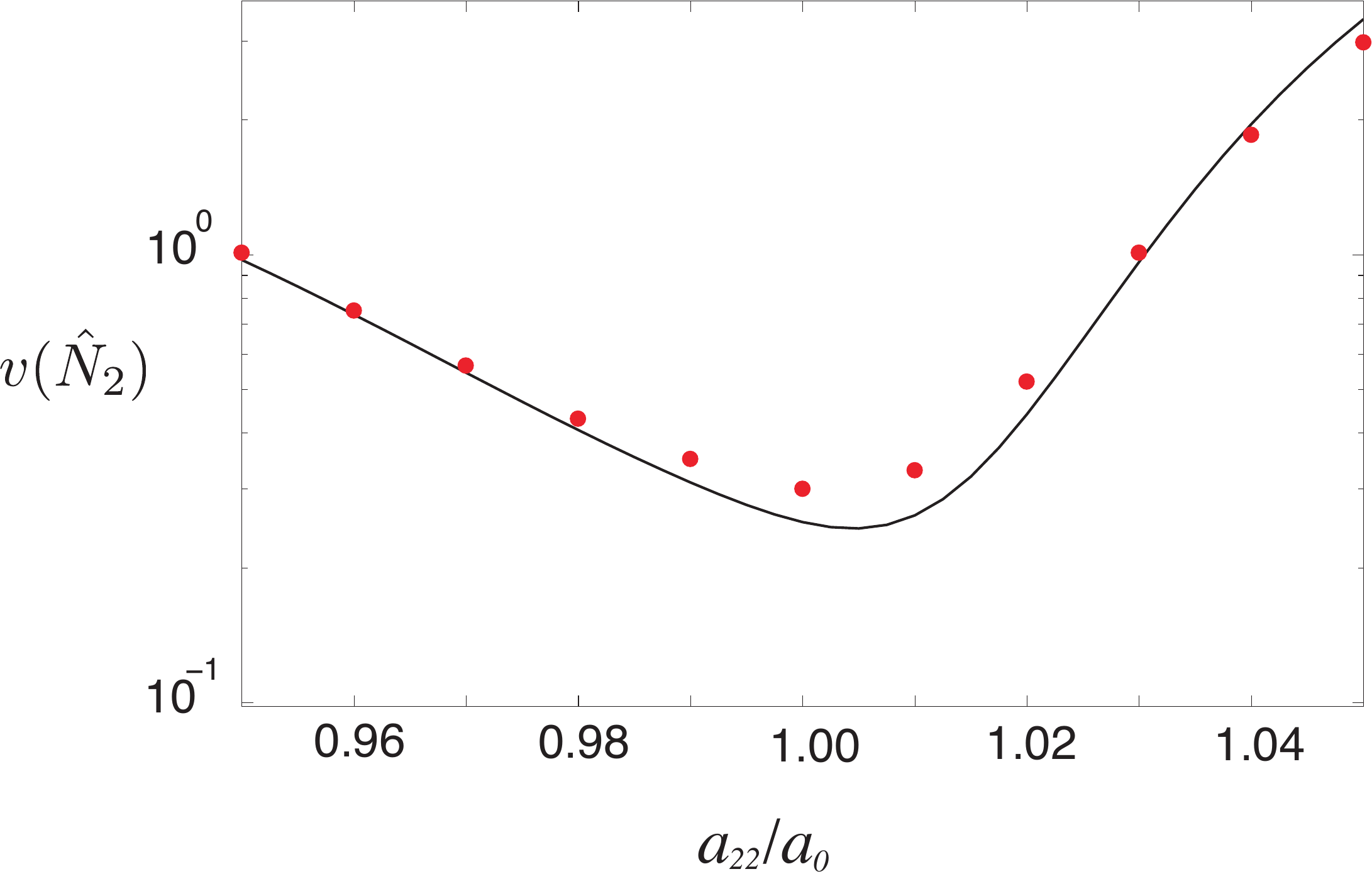}
\caption{\label{fixed_phi} (color online)  $v(\hat{N}_2)$ as $a_{22}$ is varied. All other parameters ($\Omega_0$, $\Delta t_1$, $\Delta t_2$, and $t_{\mathrm{hold}}$) are the same as in Figure (\ref{variancefig}) case I, and the phase of the second coupling pulse was fixed at $\phi = 1.4 \pi$, which was the optimum phase for Figure (\ref{variancefig}) case I. The squeezing is rapidly degraded as $a_{22}$ moves away from $a_0$. 
Black trace: results of analytic two mode model. Red dots: Results from TW simulation.  }
\end{figure} 

\begin{figure}
\includegraphics[width=0.8\columnwidth]{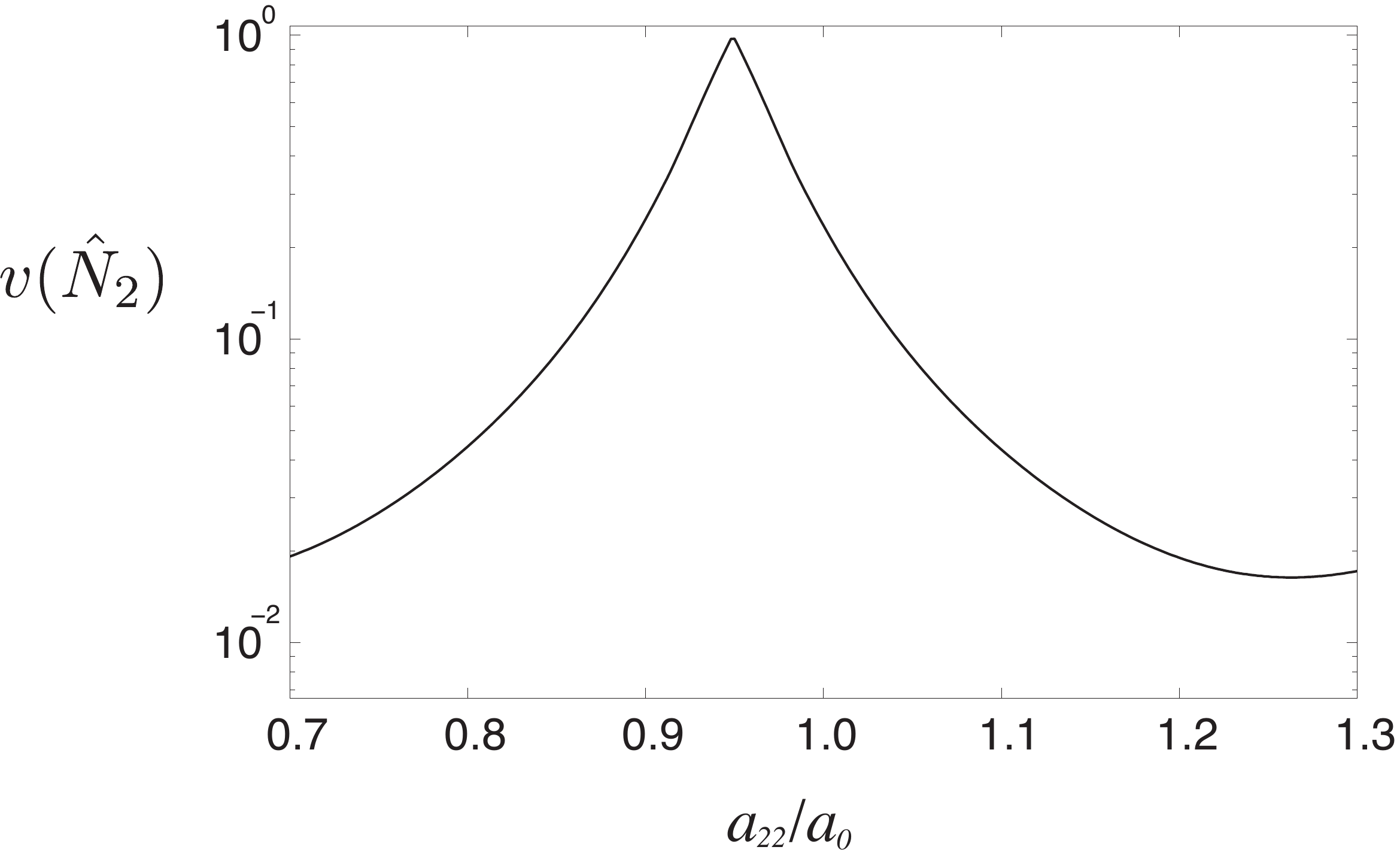}
\caption{\label{optimum_phi} $v(\hat{N}_2)$ as $a_{22}$ is varied, as calculated by the two-mode model. All other parameters ($\Omega_0$, $\Delta t_1$, $\Delta t_2$, and $t_{\mathrm{hold}}$) are the same as in Figure (\ref{variancefig}) case I. The phase of the second coupling pulse, $\phi$, has been optimized for maximum number squeezing for each value of $a_{22}$. Significant number squeezing can still be obtained as $a_{22}$ is altered by $30\%$ in either direction. The exception is as $a_{22}$ approaches $a_{11}$ and $a_{12}$ the squeezing vanishes.    }
\end{figure}

Figure (\ref{fixed_phi}) shows $v(\hat{N}_2)$ as $a_{22}$ is varied ($a_{22} = a_0$ corresponds to the value of $a_{22}$ used in Figure (\ref{variancefig}), ie $a_{0} \equiv 3.0$ nm), while keeping the phase of the second coupling pulse $\phi$ fixed at the optimum phase for squeezing ($\phi = 1.4 \pi$) as found in case I. The squeezing is completely degraded as the scatting length changes by about $3\%$. However, this is not the because the degree of phase shearing has been significantly altered. The different scattering length causes a slight shift in the mean phase between the two modes, such that it is shifted away from the optimum phase for number squeezing. If we were to re-scan the phase of the second coupling pulse to search for number squeezing (this would mean performing more shots of the experiment in order to find the optimum phase), we may find that the squeezing is still present for a large range of scattering lengths. Figure (\ref{optimum_phi}) shows  $v(\hat{N}_2)$ vs. $a_{22}$ for the same parameters as case I, but this time optimizing the phase of the second coupling pulse $\phi$ for each value of $a_{22}$. We see that significant number squeezing can still be achieved as we alter $a_{22}$ by $30\%$ in either direction. The exception is as $a_{22}$ approaches $a_{11}$ and $a_{12}$, the squeezing vanishes. 

When modeling the system with the multimode TW model, we found that squeezing could still be obtained as we varied $a_{22}$ by about $10\%$, giving decent agreement with Figure (\ref{optimum_phi}) in this range. However, we found that as $|a_{22}-a_{11}|$ became larger, the results differed significantly from the two mode model. As the TW model requires significantly more computational resources, we did not optimize the phase directly in this model. Rather, we used the optimum phase as found by the two mode model (Figure \ref{optimum_phi}). For example, with $a_{22} / a_0 = 0.7$, the two mode model predicts $v(\hat{N}_2) = 0.0192$ (at $\phi = 0.899\pi$), where as for the same value of $\phi$ the multimode TW predicts $v(\hat{N}_2) = 14.5$. It is possible that the multimode TW model predicts squeezing for some parameters for this value of the scattering length. However, as the parameter space is large ($\Delta t_1$, $\Delta t_2$, $t_{\mathrm{hold}}$, and $\phi$ can all be varied to find the optimum parameter regime), we found it almost impossible to find squeezing by searching using the multimode TW model alone. However, by utilizing the Gross-Pitaeveskii equation we were able to search for a regime where the system behaves approximately as a two-mode system, and then use the two-mode model to investigate the squeezing properties. We can then confirm the results by using the multimode TW model. We found that significant squeezing was achievable, with $v(\hat{N}_2) = 0.266$ for $\Delta t_1= 0.5$ $\mu$s, $\Delta t_2 = 0.25$ $\mu$s, $t_{\mathrm{hold}} = 16$ ms, and $\phi = 0.536 \pi$. 

Detection of atoms with high quantum efficiency will be required in order to observe the squeezing. This is experimentally challenging, but has been demonstrated before \cite{esslinger, oberthaler_nature}.  An addition effect that may degrade the squeezing is atomic loss due to inelastic collisions. Using the three-body recombination rates recently measured in \cite{ketterle_Na_dipoletrap}, we estimate that roughly $10\%$ of the atoms from state $|1\rangle$ (the state that we are not looking for squeezing) are lost during the $16$ ms hold time. However, a more important concern is the two-body inelastic collision rate, as it scales as $\int |\psi_0(\boldr)|^4 \, \dr$, that is, the same way as $\chi_{ij}$, the nonlinear interaction parameter, so reducing the atomic density will not help, as the lifetime due to collisions and the time taken to achieve squeezing scale identically. It was observed in \cite{ketterle_Na_dipoletrap} that with mixtures of different hyperfine states decayed on timescales of order several milliseconds. However, if the maximal stretched combination of states was used \cite{footnote1} (for example $|F = 1, m_F = 1\rangle$, $|F=2, m_F = 2\rangle$), lifetimes of several seconds were observed. Using this particular combination of states would allow for squeezing under our scheme, as it has $a_{11} \neq a_{12}$, and as these scattering lengths are similar one to the ones used in this paper, one would expect a similar amount of squeezing.

It may be possible to create an intensity squeezed atom laser via a similar technique to that discussed in this Letter. By outcoupling two co-propagating hyperfine states, and interfering them at particular distance from the condensate, it may be possible to create intensity squeezing in one of the modes. However, this would require a species of atom with two magnetic field insensitive states, with scattering lengths $a_{11} + a_{22} \neq 2a_{12}$. This requirement could be avoided by using a separated beam path interferometer, as the effective $\chi_{12}$ goes to zero. However, it may be difficult to achieve the required mode matching in this case. 

Finally, we wish to note that the specific states used here are illustrative, not optimal. We have presented a scheme that allows the the generation of significant amounts of number squeezing in a BEC, and demonstrated its effectiveness for plausible states of an atom that can be Bose-condensed. However scattering lengths are not well-known for various states of many atomic species, and better candidates for generating number squeezed BECs almost certainly exist.

\section{Acknowledgments}
The authors would like to acknowledge useful discussions with Andy Ferris, and support from the Australian NCI supercomputing facility. This work was supported by the Australian Research Council Centre of Excellence for Quantum-Atom Optics, and by Australian Research Council discovery project DP0986893.


\begin{thebibliography}{99}
\bibitem{footnote1} We were unaware of the favorable inelastic loss rates for the $|F = 1, m_F = 1\rangle$, $|F=2, m_F = 2\rangle$ collisions at the time when we performed the calculations.
\end{thebibliography}
\end{document}